\documentclass[prd,onecolumn,nofootinbib]{revtex4}
\usepackage{amsmath,amssymb}

\begin{document}

\title{Spacetime torsion as a possible remedy to major problems in gravity and cosmology}
\author{Nikodem J. Pop{\l}awski}
\affiliation{Department of Physics, Indiana University, Swain Hall West, 727 East Third Street, Bloomington, Indiana 47405, USA}
\email{nipoplaw@indiana.edu}
\date{\today}

\begin{abstract}
We show that the Einstein-Cartan-Sciama-Kibble theory of gravity with torsion not only extends general relativity to account for the intrinsic spin of matter, but it may also eliminate major problems in gravitational physics and answer major questions in cosmology.
These problems and questions include: the origin of the Universe, the existence of singularities in black holes, the nature of inflation and dark energy, the origin of the matter-antimatter asymmetry in the Universe, and the nature of dark matter.
\end{abstract}

\maketitle

The big-bang cosmology, based on Einstein's general theory of relativity (GR), successfully describes primordial nucleosynthesis and predicts the cosmic microwave background radiation.
It does not, however, address some fundamental questions.
What is the origin of the rapid expansion of the Universe from an initial, extremely hot and dense state?
What is the nature of dark energy?
And what caused the observed asymmetry between matter and antimatter in the Universe?
The validity of GR also breaks inside black holes and at the beginning of the big bang, where the matter is predicted to compress indefinitely to curvature singularities.
The big-bang cosmology also requires an inflationary scenario with additional fields to explain why the present Universe appears spatially flat, homogeneous and isotropic.
In this essay, we briefly describe how the solution to the above questions and problems may come from an old adaptation of GR, the Einstein-Cartan-Sciama-Kibble (ECSK) theory of gravity \cite{Kib,torsion,Lo}.
This classical theory naturally extends GR to account for the quantum-mechanical, intrinsic angular momentum (spin) of elementary particles that compose gravitating matter, causing spacetime to exhibit a geometric property called torsion.

The ECKS gravity is based on the gravitational Lagrangian density proportional to the curvature scalar, as in GR.
This theory, however, removes the GR restriction of the affine connection $\Gamma^{\,\,k}_{i\,j}$ be symmetric.
Instead, the antisymmetric part of the connection $S^k_{\phantom{k}ij}=\Gamma^{\,\,\,\,k}_{[i\,j]}$ (torsion tensor) is regarded as a dynamical variable like the metric tensor $g_{ik}$.
Varying the total action for the gravitational field and matter with respect to the metric gives the Einstein equations that relate the curvature to the dynamical energy-momentum tensor $T_{ik}=(2/\sqrt{-g})\delta\mathfrak{L}/\delta g^{ik}$, where $\mathfrak{L}$ is the matter Lagrangian density and $g=\mbox{det}(g_{ik})$ \cite{LL}.
These equations can be written in a GR form as $G_{ik}=\kappa(T_{ik}+U_{ik})$, where the source is the modified energy-momentum tensor with an additional term $U_{ik}$ quadratic in the torsion tensor.
$G_{ik}$ is the standard Einstein tensor and $\kappa=8\pi G/c^4$.
These equations can also be written as $R_{ik}-R^j_j g_{ik}/2=\kappa\theta_{ik}$, where $R_{ik}(\Gamma)$ is the Ricci tensor of the connection and $\theta_{ik}$ is the canonical energy-momentum tensor.

Varying the total action with respect to the torsion gives the Cartan equations \cite{Kib,torsion,Lo},
\begin{equation}
S_{jik}-S^l_{\phantom{l}il}g_{jk}+S^l_{\phantom{l}kl}g_{ji}=-\frac{1}{2}\kappa s_{ikj}.
\label{Car}
\end{equation}
These equations relate the torsion tensor to the spin tensor of matter $s_{ij}^{\phantom{ij}k}=(2/\sqrt{-g})\delta\mathfrak{L}/\delta C^{ij}_{\phantom{ij}k}$, where $C^{ij}_{\phantom{ij}k}$ is the contortion tensor.
Since they are linear and algebraic, the torsion tensor vanishes outside material bodies where the spin density is zero.

Upon substituting (\ref{Car}), the tensor $U_{ik}$ becomes \cite{Kib,torsion}
\begin{equation}
U^{ik}=\frac{1}{2}\kappa\biggl(s^{ij}_{\phantom{ij}j}s^{kl}_{\phantom{kl}l}-s^{ij}_{\phantom{ij}l}s^{kl}_{\phantom{kl}j}-s^{ijl}s^k_{\phantom{k}jl}+\frac{1}{2}s^{jli}s_{jl}^{\phantom{jl}k}+\frac{1}{4}g^{ik}(2s^{\phantom{j}l}_{j\phantom{l}m}s^{jm}_{\phantom{jm}l}-2s^{\phantom{j}l}_{j\phantom{l}l}s^{jm}_{\phantom{jm}m}+s^{jlm}s_{jlm})\biggr).
\label{Ein}
\end{equation}
The contribution to the curvature from the spin is proportional to $\kappa^2$, so it is significant only at densities of matter much larger than the nuclear density.
Such extremely high densities existed in the very early Universe and exist inside black holes.
In other physical situations, the ECSK gravity effectively reduces to GR, passing its experimental and observational tests.

Quarks and leptons, which carry the $1/2$ intrinsic spin, are the most relevant source of torsion because the torsion tensor appears in the Dirac Lagrangian via the covariant derivative of a spinor with respect to the affine connection \cite{Kib,torsion,Lo}.
The spin tensor for a Dirac spinor $\psi$ is totally antisymmetric and it is given by \cite{Kib,torsion,Lo}
\begin{equation}
s^{ijk}=\frac{i\hbar c}{2}\bar{\psi}\gamma^{[i}\gamma^j\gamma^{k]}\psi,
\label{spin}
\end{equation}
where $\bar{\psi}$ is the Dirac conjugate spinor and $\gamma^i$ are the Dirac matrices $\gamma^{(i}\gamma^{j)}=2g^{ij}$.
Because of the Cartan equations (\ref{Car}), the torsion tensor is quadratic in spinor fields.
Its substitution into the Dirac equation gives the cubic Hehl-Datta equation for $\psi$ \cite{Dirac,HI}.
\begin{equation}
i\hbar\gamma^k\psi_{:k}=mc\psi-\frac{3\kappa c\hbar^2}{8}(\bar{\psi}\gamma^5\gamma_k\psi)\gamma^5\gamma^k\psi,
\label{HD}
\end{equation}
where the colon denotes a GR covariant derivative (with respect to the Christoffel symbols) and $m$ is the mass of the fermion.
For a spinor with electric charge $q$ in the presence of the electromagnetic potential $A_k$, $\psi_{:k}$ is generalized to $\psi_{:k}-iqA_k\psi/(\hbar c)$, so that (\ref{HD}) becomes
\begin{equation}
i\hbar\gamma^k\psi_{:k}+\frac{q}{c}A_k\gamma^k\psi=mc\psi-\frac{3\kappa c\hbar^2}{8}(\bar{\psi}\gamma^5\gamma_k\psi)\gamma^5\gamma^k\psi.
\label{HD_em}
\end{equation}

The second term on the right of (\ref{HD}) corresponds to an effective axial-axial, four-fermion interaction \cite{Kib,Dirac},
\begin{equation}
\mathfrak{L}_{\textrm{e}}=\frac{3\kappa c\hbar^2\sqrt{-g}}{16}(\bar{\psi}\gamma^5\gamma_k\psi)(\bar{\psi}\gamma^5\gamma^k\psi).
\label{four}
\end{equation}
Although such an interaction term appears nonrenormalizable, $\mathfrak{L}_\textrm{e}$ is an effective Lagrangian density in which only the metric tensor and spinor fields are dynamical variables.
The original Lagrangian density for a Dirac field, in which the torsion tensor is also a dynamical variable (leading to the Hehl-Datta equation), is quadratic in spinor fields and hence renormalizable.

At macroscopic scales, the contributions to torsion from fermions must be averaged.
Applying the Papapetrou-Nomura-Shirafuji-Hayashi method of multipole expansion in the Riemann-Cartan spacetime \cite{NSH} to the conservation law for the spin density, $s^{ijk}_{\phantom{ijk},k}-\Gamma^{\,\,i}_{l\,k}s^{jlk}+\Gamma^{\,\,j}_{l\,k}s^{ilk}-2\theta^{[ij]}=0$ (which results from the Bianchi identities in the ECSK gravity \cite{Kib,torsion,Lo}), yields in the point-particle approximation
\begin{equation}
s_{ij}^{\phantom{ij}k}=s_{ij}u^k,\,\,\,s_{ij}u^j=0,
\label{fluid}
\end{equation}
where $u^i$ is the four-velocity of the macroscopic matter and $s_{ij}$ is a tensor.
The relations (\ref{fluid}) describe a Weyssenhoff spin fluid \cite{spin_fluid}.
The macroscopic canonical energy-momentum tensor of a spin fluid is
\begin{equation}
\theta_{ij}=c\Pi_i u_j-p(g_{ij}-u_i u_j),
\label{enmo}
\end{equation}
where $\Pi_i$ is the four-momentum density and $p$ is its pressure.

Even if the spin orientation of particles is random, the macroscopic spacetime average of terms that are quadratic in the spin tensor does not vanish.
In this case, the Einstein equations give \cite{avert_avg,Gas}
\begin{equation}
G^{ij}=\kappa\Bigl(\epsilon-\frac{1}{4}\kappa s^2\Bigr)u^i u^j-\kappa\Bigl(p-\frac{1}{4}\kappa s^2\Bigr)(g^{ij}-u^i u^j),
\label{Einst}
\end{equation}
where $\epsilon=c\Pi_i u^i$ is the rest energy density and
\begin{equation}
s^2=s_{ij}s^{ij}/2=(\hbar cn)^2/8>0
\end{equation}
is the average square of the spin density ($n$ being the fermion number density) \cite{spin_den}.
Thus the Einstein-Cartan equations for a spin fluid are equivalent to the Einstein equations for a perfect fluid with the effective energy density $\epsilon-\kappa s^2/4$ and the effective pressure $p-\kappa s^2/4$.

The Friedman equations for a closed Friedman-Lema\^{i}tre-Robertson-Walker (FLRW) universe filled with a spin-fluid relativistic matter are given by \cite{Gas,Kuc}
\begin{eqnarray}
& & {\dot{a}}^2+1=\frac{1}{3}\kappa\Bigl(\epsilon-\frac{1}{4}\kappa s^2\Bigr)a^2, \label{Fri1} \\
& & \frac{d}{dt}\bigl((\epsilon-\kappa s^2/4)a^3\bigr)+(p-\kappa s^2/4)\frac{d}{dt}(a^3)=0,
\label{Fri2}
\end{eqnarray}
where $a$ is the cosmic scale factor.
The relative sign between $\epsilon$ and $\kappa s^2/4$ in the effective energy density is negative.
Accordingly, torsion manifests itself as a force that counters gravitational attraction.
The conservation law (\ref{Fri2}) implies that $s^2$ scales with the cosmic scale factor as $\sim a^{-6}$, whereas $\epsilon$ (in the early, ultrarelativistic Universe) scales as $\sim a^{-4}$.
Torsion therefore prevents the collapsing spin-fluid matter from reaching a singularity \cite{Gas,Kuc} (the avoidance of singularities for matter composed of oriented spins has been shown in \cite{avert}).
Instead, the Universe has a minimum but finite scale factor at which $\kappa s^2/4=\epsilon$.
Accordingly, the singular big bang is replaced by a nonsingular big bounce from a contracting universe \cite{bounce}.
This result agrees with Hawking-Penrose singularity theorems because the contribution from torsion violates energy conditions at high densities \cite{avert_avg}.
Interestingly, loop quantum gravity also predicts a cosmic bounce \cite{LQG}.

The Friedman equation (\ref{Fri1}) can be written as the evolution equation for the total density parameter $\Omega$ \cite{infl}:
\begin{equation}
\Omega=1+\frac{(\Omega_0-1)\hat{a}^4}{\Omega_{R0}\hat{a}^2+\Omega_{S0}},
\label{omega}
\end{equation}
where $\hat{a}=a/a_0$ ($a_0$ being the present scale factor), $\Omega_0$ is the present-day total density parameter, $\Omega_{R0}$ is the present-day radiation density parameter, and $\Omega_{S0}=-\kappa s^2_0/(4\epsilon_c)$ is the present-day torsion density parameter ($s^2_0$ being the present value of $s^2$ and $\epsilon_c/c^2$ being the present critical density).
In GR, $\Omega_{S0}$ does not appear in (\ref{omega}), so $\Omega$ tends to 1 as $a\rightarrow 0$, introducing the flatness problem in big-bang cosmology because $\Omega$ at the GUT epoch must have been tuned to 1 to a precision of more than 52 decimal places in order for $\Omega$ to be near 1 today.
The horizon problem is related to the above flatness problem.

In the ECSK gravity, the function (\ref{omega}) tends to infinity as $a$ tends to its minimum value, and has a local minimum where it is equal to $1-4\Omega_{S0}(\Omega_0-1)/\Omega^2_{R0}\approx1+10^{-63}$.
As the Universe expands, $\Omega$ rapidly decreases from infinity to the local minimum and then increases according to the standard dynamics.
Thus the apparent fine tuning of $\Omega$ in the very early Universe is naturally caused by the extremely small in magnitude, negative torsion density parameter $\Omega_S$, solving the flatness problem without cosmic inflation \cite{infl}.
The velocity of the point that is antipodal to the coordinate origin is $\dot{a}=\pi (\Omega-1)^{-1/2}\,c$.
At the point where $\Omega$ has a minimum, such a velocity has a maximum where it is $\pi\Omega_{R0}(-\Omega_{S0}(\Omega_0-1))^{-1/2}/2\,c\approx 10^{32}\,c$.
Therefore, the value of $\Omega_{S0}$ causes that the Universe, expanding from its minimum size, forms approximately $10^{32}$ causally disconnected volumes from a single causally connected region of spacetime, solving the horizon problem without cosmic inflation \cite{infl}.

Before the bounce at the minimum size, the Universe was contracting.
Since the matter inside a black hole, instead of compressing to a singularity, rebounds and begins to expand, a natural scenario follows according to which every black hole produces a new, closed nonsingular universe inside \cite{BH}.
Extremely strong gravitational fields near the bounce cause an intense pair production, which generates the observed amount of mass and increases the energy density \cite{part}.
Such a particle production does not change the total (matter plus gravitational field) energy of the resulting FLRW universe \cite{energy}.
The equation of state of the matter remains stiff, $\epsilon=p$, until the bounce \cite{stiff_mat,stiff_cosm}.
After the bounce, the matter begins to expand as a new universe with mass \cite{mass}
\begin{equation}
M_{\textrm{univ}}\approx\frac{M^2_\textrm{BH} m_\textrm{n}}{m_\textrm{Pl}^2},
\end{equation}
where $m_\textrm{n}$ is the mass of a neutron, $m_\textrm{Pl}$ is the Planck mass, and $M_\textrm{BH}$ is the mass of the black hole for observers outside it.
Such an expansion is not visible for those observers, in whose frame of reference the horizon's formation and all subsequent processes occur after infinite time.
The new universe is thus a separate spacetime branch with its own timeline.
Such a universe can grow infinitely large if the cosmological constant $\Lambda$ is present and if its total mass $M_{\textrm{univ}}>c^2/(3G\sqrt{\Lambda})$ \cite{Lo,dyn}, otherwise it would expand to a finite maximum size and then contract to the next bounce \cite{Kuc}.
Equaling $M_{\textrm{univ}}$ to the mass of our Universe, which obeys the above inequality, gives $M_{\textrm{BH}}\sim 10^3\,M_\odot$.
The Universe may therefore have originated from the interior of an intermediate-mass black hole.

Although the ECKS gravity are time-symmetric, the boundary conditions of a universe in a black hole are not, because the motion of matter through the event horizon of a black hole is unidirectional.
The time-asymmetric collapse of matter through the event horizon therefore defines the arrow of time of such a universe before the subsequent expansion \cite{infl}.

Torsion, via the Hehl-Datta equation (\ref{HD}), may also explain the nature of dark energy, the observed matter-antimatter imbalance in the Universe and the origin of dark matter.
If fermion fields have a nonzero vacuum expectation value than the corresponding four-fermion interaction term (\ref{four}) acts like a cosmological constant \cite{dark},
\begin{equation}
\Lambda=(3\kappa^2 c\hbar^2/16)\langle0|(\bar{\psi}\gamma^5\gamma_k\psi)(\bar{\psi}\gamma^5\gamma^k\psi)|0\rangle.
\end{equation}
For condensing quarks, such a torsion-induced cosmological constant is positive and its energy scale is only about 8 times larger than the observed value \cite{dark,Zel}:
\begin{equation}
\Lambda=\frac{\kappa^2 c\hbar^2}{3}\bigl(\langle0|\bar{\psi}\psi|0\rangle\bigr)^2\propto\frac{\lambda^6_{\textrm{QCD}}}{m^2_{\textrm{Pl}}},
\end{equation}
where $\lambda_{\textrm{QCD}}$ is the QCD scale parameter of the SU(3) gauge coupling constant.
The classical Hehl-Datta equation also turns out to be asymmetric under a charge-conjugation transformation.
For the charge-conjugate spinor field $\psi^c=-i\gamma^2\psi^\ast$, (\ref{HD_em}) becomes \cite{bar}:
\begin{equation}
i\hbar\gamma^k\psi_{:k}^c-\frac{q}{c}A_k\gamma^k\psi^c=mc\psi^c+\frac{3\kappa c\hbar^2}{8}(\overline{\psi^c}\gamma^5\gamma_k\psi^c)\gamma^5\gamma^k\psi^c.
\end{equation}
A classical Dirac spinor and its charge-conjugate therefore satisfy different field equations, which leads to their decay asymmetry at extremely high densities in the early Universe \cite{bar}.
This asymmetry may be responsible for a scenario, according to which dark matter is composed of antimatter \cite{baryo}.

Lastly, quantum field theory based on the Hehl-Datta equation may avert divergent integrals in calculating radiative corrections.
It has been shown in \cite{Mer} that a nonlinear spinor theory resulting from a cubic equation $i\hbar\gamma^k\psi_{,k}=mc\psi\pm l^2(\bar{\psi}\psi)\psi$, where $l^2=\mbox{const}$, may exhibit self-regulation of its ultraviolet behavior.
A similar propagator self-regulation should therefore occur for the Hehl-Datta equation.
In addition, the multipole expansion applied to Dirac fields in the ECSK gravity shows that such fields cannot form singular, point-like configurations because these configurations would violate the conservation law for the spin density and thus the Bianchi identities \cite{non}.
Instead, they describe nonsingular particles whose spatial dimensions are at least on the order of their Cartan radii $r_C$, defined by a condition $\epsilon\sim\kappa s^2$ \cite{Tr}, where $\epsilon\sim mc^2|\psi|^2$, $\sqrt{s^2}\sim \hbar c|\psi|^2$, and the wave function $\psi\sim r^{-3/2}_C$ \cite{non}.
Accordingly, the de Broglie energy associated with the Cartan radius of a fermion ($\sim10^{-27}\,$m for an electron) should introduce an effective ultraviolet cutoff for such a fermion in quantum field theory in the ECKS spacetime \cite{non}.
The avoidance of divergences in radiative corrections in quantum field theory may thus come from the same mechanism that can prevent the formation of singularities from matter composed of quarks and leptons: from spacetime torsion coupled to intrinsic spin.

As a final remark, we note that although there exist many other theories of gravity with torsion (\cite{rev} and references therein), the ECKS gravity is the closest one to GR.
The teleparallel theory of gravity (\cite{tel} and references therein), with nonzero torsion and vanishing curvature, is merely another (and less natural with regard to the choice of the gravitational Lagrangian) formulation of GR.
In fact, the ECSK theory should not be regarded as an alternative to GR, but rather as its natural extension (to sources with intrinsic spin) because it has the same simple Lagrangian, proportional to the Ricci scalar.
In this theory, the torsion tensor is a dynamical variable without any constraints on its form.
As a result, the spin of matter is a source of torsion, which has a natural physical interpretation in the context of the Poincar\'{e} group \cite{Kib,torsion}.
Furthermore, the ECSK gravity predicts that the torsion tensor is related to the spin density through a linear, algebraic equation, so that torsion does not propagate and vanishes where matter is absent \cite{Kib,torsion,rev}.
Therefore, in this theory, the principle of equivalence is satisfied where it should be: in empty space.
Another advantage of the ECSK theory relative to other theories with torsion (and almost all modified theories of gravity) is that it has no free parameters, like GR.

\end{document}